\documentstyle[preprint,aps,floats]{revtex}
\ifx\nopictures Y \else \input epsf.tex \fi
\begin{document}
\draft
\tighten
\baselineskip=18pt

\title{%
Testing Multiple Gluon Dynamics at the Tevatron}
\author{%
C. Bal\'azs\thanks{Corresponding author. E-mail address: balazs@pa.msu.edu} and 
C.--P. Yuan\thanks{E-mail address: yuan@pa.msu.edu}}
\date{\today}
\address{%
Department of Physics and Astronomy, Michigan State University, \\
East Lansing, MI 48824, U.S.A.}
\maketitle

\begin{flushleft}
{MSUHEP-70317 \\ CTEQ-703}
\end{flushleft}

\begin{abstract}
We propose the measurement of the ratio $R_{CSS}(Q_T^{\min})\equiv \frac{\sigma
(Q_T>Q_T^{\min })}{\sigma _{Total}}$ to study the effects of the multiple
soft gluon radiation, predicted by QCD, on the transverse momentum ($Q_T$)
distribution of the weak gauge bosons $W^{\pm }$ and $Z^0$ produced at the
Tevatron. We compare the prediction of the extended Collins-Soper-Sterman
resummation formalism with the next-to-leading and
next-to-next-to-leading order calculations. We show that both the rich
dynamics of the QCD multiple soft gluon radiation and the 
non-perturbative sector of QCD can be tested by measuring $R_{CSS}$.
\end{abstract}


\section{Introduction}

With $100\,{\rm pb}^{-1}$ luminosity at the Tevatron, about $2\times 10^6$ 
$W^{\pm }$ and $6\times 10^5$ $Z^0$ bosons are produced, and the data sample
will increase by a factor of 20 in the Run 2 era. In view of this large
event rate, a careful study of the transverse momentum distributions of
vector bosons can provide a stringent test of the rich dynamics of the
multiple soft gluon emission predicted by Quantum Chromodynamics (QCD). 
The increasing 
precision of the experimental data demands a high precision theoretical
calculation of the distributions of the $W^{\pm}$ and $Z^0$ bosons, which
takes the effects of multiple gluon radiation into account. 
In this work, within the extended
Collins-Soper-Sterman formalism ~\cite{CSS,Balazs-Qui-Yuan,Balazs-Yuan}, 
we illustrate the effect of the multiple gluon radiation on the vector boson 
transverse momentum distribution.

To test the dynamics of the multiple soft gluon radiation, 
in this work we propose the measurement of the ratio 
$R_{CSS}(Q_T^{\min})\equiv \frac{\sigma (Q_T>Q_T^{\min })}{\sigma _{Total}}$ 
for the $W^{\pm }$ and $Z^0$
bosons produced at the Tevatron. We show that for vector boson transverse
momenta less than about 30 GeV, the difference between the
resummed and the fixed order predictions (either at the $\alpha _S$ or 
$\alpha _S^2$ order) can be observed by measurement. This suggests that in
this kinematic region, the effects of the multiple soft
gluon radiation are important, and hence, this case provides a ideal
opportunity to test this aspect of the QCD dynamics. For $Q_T$ less than
about 10 GeV, the $Q_T$ distribution is largely determined by the
non-perturbative part of QCD. At the Tevatron, for $W^{\pm }$
and $Z^0$ production, this non-perturbative physics, when parametrized by
Eq.~(\ref{eq:WNPLY}), is dominated by the parameter $g_2$, which was shown to
be related to properties of the QCD vacuum~\cite{Korchemsky-Sterman}.
Therefore, precisely measuring the $Q_T$ distributions in the low $Q_T$
region, e.g. from $Z^0$ events can advance our knowledge of the
non-perturbative QCD physics.

The next section summarizes the relevant formulae of our extension of the  
Collins-Soper-Sterman resummation formalism, which describes the production 
and decay of vector bosons at hadron colliders. In Section \ref{sec:RCSS}, 
we present the differences in the next-to-leading order 
(NLO, e.g. ${\cal O}(\alpha_S)$), next-to-next-to-leading order
(NNLO, e.g. ${\cal O}(\alpha_S^2)$), and the resummed predictions for 
$R_{CSS}$, and show that the differences due to the soft gluon effects are 
measurable. 
We illustrate that, using the experimental data, the improvement of the 
non-perturbative sector of the resummation formalism is also possible. 
Finally, we draw our conclusions, based upon these theoretical results, 
about the importance of the multiple soft gluon radiation in $W^\pm$ and $Z^0$ 
production at the Tevatron.

\section{Summary of the Resummation Formalism}

The fully differential cross section of the hadronic production and decay of
a vector boson, within the extended Collins-Soper-Sterman resummation
formalism, is characterized as follows. The kinematics of
the vector boson $V$ (real or virtual) can be expressed in the terms of its
mass $Q$, rapidity $y$, transverse momentum $Q_T$, and azimuthal angle $\phi
_V$, measured in the laboratory frame (the center-of-mass frame of hadrons 
$h_1$ and $h_2$). The kinematics of the lepton $\ell _1$ is described by 
$\theta $ and $\phi $, the polar and the azimuthal angles defined in the
Collins-Soper frame~\cite{CSFrame}, which is a special rest frame of the 
$V$-boson~\cite{lamtung}. (A more detailed discussion of the kinematics can
be found in Ref.~\cite{Balazs-Yuan}.) The resummed cross section is given by
the following formula in Ref.~\cite{Balazs-Qui-Yuan}: 
\begin{eqnarray}
&&\left( {\frac{d\sigma (h_1h_2\rightarrow V(\rightarrow \ell _1{\bar{\ell}_2%
})X)}{dQ^2\,dy\,dQ_T^2\,d\phi _V\,d\cos {\theta }\,d\phi }}\right) _{res}={%
\frac 1{96\pi ^2S}}\,{\frac{Q^2}{(Q^2-M_V^2)^2+Q^4\Gamma _V^2/M_V^2}} 
\nonumber \\
&&~~\times \left\{ {\frac 1{(2\pi )^2}}\int d^2b\,e^{i{\vec{Q}_T}\cdot {\vec{%
b}}}\,\sum_{j,k}{\widetilde{W}_{j{\bar{k}}}(b_{*},Q,x_1,x_2,\theta ,\phi
,C_1,C_2,C_3)}\,\widetilde{W}_{j{\bar{k}}}^{NP}(b,Q,x_1,x_2)\right. 
\nonumber \\
&&~~~~\left. +~Y(Q_T,Q,x_1,x_2,\theta ,\phi ,{C_4})\right\} .
\label{eq:ResFor}
\end{eqnarray}
In the above equation  
the parton momentum fractions are defined as $x_1=e^yQ/\sqrt{S}$ and 
$x_2=e^{-y}Q/\sqrt{S}$, where $\sqrt{S}$ is the center-of-mass (CM) energy of
the hadrons $h_1$ and $h_2$. For $V=W^{\pm }$ or $Z^0$, we adopt the
LEP line-shape prescription of the resonance behavior. The renormalization
group invariant quantity $\widetilde{W}_{j{\bar{k}}}(b)$, which sums to all
orders in $\alpha _S$ all the singular terms that behave as 
$\alpha_S^n Q_T^{-2} \ln ^{2m -1}{(Q_T^2/Q^2)}$ ($1 \leq m \leq n$) 
for $Q_T\rightarrow 0$, is 
\begin{eqnarray}
&&\widetilde{W}_{j{\bar{k}}}(b,Q,x_1,x_2,\theta ,\phi {,C_1,C_2,C_3})=\exp
\left\{ -{\cal S}(b,Q{,C_1,C_2})\right\} \mid V_{jk}\mid ^2  \nonumber \\
&&~\times \left\{ \left[ \left( C_{ja}\otimes f_{a/h_1}\right) (x_1)~\left(
C_{{\bar{k}}b}\otimes f_{b/h_2}\right) (x_2)+\left( C_{{\bar{k}}a}\otimes
f_{a/h_1}\right) (x_1)~\left( C_{jb}\otimes f_{b/h_2}\right) (x_2)\right]
\right.  \nonumber \\
&&~~~~~\times (g_L^2+g_R^2)(f_L^2+f_R^2)(1+\cos ^2\theta )  \nonumber \\
&&~~~+\left[ \left( C_{ja}\otimes f_{a/h_1}\right) (x_1)~\left( C_{{\bar{k}}%
b}\otimes f_{b/h_2}\right) (x_2)-\left( C_{{\bar{k}}a}\otimes
f_{a/h_1}\right) (x_1)~\left( C_{jb}\otimes f_{b/h_2}\right) (x_2)\right] 
\nonumber \\
&&~~~~~\left. \times (g_L^2-g_R^2)(f_L^2-f_R^2)(2\cos \theta )\right\} ,
\label{eq:WTwi}
\end{eqnarray}
where $\otimes $ denotes the convolution 
\[
\left( C_{ja}\otimes f_{a/h_1}\right) (x_1)=\int_{x_1}^1{\frac{d\xi _1}{\xi
_1}}\,C_{ja}\left( {\frac{x_1}{\xi _1}},b,\mu =\frac{C_3}b,C_1,C_2\right)
\;f_{a/h_1}\left( \xi _1,\mu =\frac{C_3}b\right) , 
\]
and the $V_{jk}$ coefficients are given by 
\begin{eqnarray*}
V_{jk}=\cases{ {\rm Cabibbo-Kobayashi-Maskawa~matrix~elements} & for $V =
W^\pm $ \cr $$\delta_{jk}$$ & for $V = Z^0$}.
\end{eqnarray*}
The $q{\bar q^{\prime}}V$ and $\ell_1 {\bar \ell_2} V$ vertices are defined as 
$i \gamma_\mu [g_L (1 - \gamma_5) + g_R (1 + \gamma_5)]$ and 
$i \gamma_\mu [f_L (1 - \gamma_5) + f_R (1 + \gamma_5)]$, respectively.
For example, for $V=W^{+}$, $q=u$, ${\bar q^{\prime }}={\bar d}$, $\ell_1=\nu
_e $, and ${\bar \ell_2}=e^{+}$, the couplings are 
$g_L^2=f_L^2=G_FM_W^2/\sqrt{2}$ and 
$g_R^2=f_R^2=0$, where $G_F$ is the Fermi constant.
The Sudakov exponent ${\cal S}(b,Q{,C_1,C_2})$ in Eq.~(\ref{eq:WTwi}) is
defined as 
\begin{eqnarray*}
{\cal S}(b,Q{,C_1,C_2})= \int_{C_1^2/b^2}^{C_2^2Q^2}{\frac{d{\bar \mu }^2}{{\bar \mu
}^2}} \left[ A\left( \alpha _S({\bar \mu }),C_1 \right) \ln \left( {\frac{%
C_2^2Q^2}{{\bar \mu }^2}}\right) + B\left( \alpha _S({\bar \mu }%
),C_1,C_2\right) \right] .  
\end{eqnarray*}
The explicit forms of the $A$, $B $ and $C$ functions and the renormalization
constants $C_i$ ($i$=1,2,3) are summarized in 
Refs.\cite{CSS,Balazs-Qui-Yuan}.

In Eq.~(\ref{eq:ResFor}) the magnitude of the impact parameter $b$ is
integrated from 0 to $\infty $. However, in the region where $b\gg 1/\Lambda
_{QCD}$, the Sudakov exponent ${\cal S}(b,Q,C_1,C_2)$ diverges as the result
of the Landau pole of the QCD coupling $\alpha _S(\mu )$ at $\mu =\Lambda
_{QCD}$, and the perturbative calculation is no longer reliable.
In this region of the impact parameter
space (i.e. large $b$), a prescription for parametrizing the
non-perturbative physics in the low $Q_T$ region is necessary. 
Following the idea of Collins and Soper~\cite{Collins}, 
the renormalization group invariant quantity 
$\widetilde{W}_{j{\bar{k}}}(b)$ is written as 
\[
\widetilde{W}_{j{\bar{k}}}(b)=
\widetilde{W}_{j{\bar{k}}}(b_{*})\widetilde{W}
_{j{\bar{k}}}^{NP}(b)\,.
\]
Here $\widetilde{W}_{j{\bar{k}}}(b_{*})$ is the perturbative part of 
$\widetilde{W}_{j{\bar{k}}}(b)$ and can be reliably calculated by
perturbative expansions, while $\widetilde{W}_{j{\bar{k}}}^{NP}(b)$ is the
non-perturbative part of $\widetilde{W}_{j{\bar{k}}}(b)$ that cannot be
calculated by perturbative methods and has to be determined from experimental
data. To test this assumption, one should verify that there exists a
universal functional form for this non-perturbative function $\widetilde{W}%
_{j{\bar{k}}}^{NP}(b)$. This is similar to the general expectation that
there exists a universal set of parton distribution functions (PDF's) that can
be used in any perturbative QCD calculation to compare it with experimental
data. In the perturbative part of $\widetilde{W}_{j{\bar{k}}}(b)$, 
\[
b_{*}={\frac b{\sqrt{1+(b/b_{max})^2}}}\,, 
\]
and the non-perturbative function was parametrized by (cf. Ref.~\cite{CSS}) 
\begin{equation}
\widetilde{W}_{j\bar{k}}^{NP}(b,Q,Q_0,x_1,x_2)=\exp \left[ -F_1(b)\ln \left( 
\frac{Q^2}{Q_0^2}\right) -F_{j/{h_1}}(x_1,b)-F_{{\bar{k}}/{h_2}%
}(x_2,b)\right] ,  \label{eq:Wnonpert}
\end{equation}
where $F_1$, $F_{j/{h_1}}$ and $h_{{\bar{k}}/{h_2}}$ have to be first
determined using some sets of data, and later can be used to predict the other
sets of data to test the dynamics of multiple gluon radiation predicted by
this model of the QCD theory calculation. As noted in Ref.~\cite{CSS}, $F_1$
does not depend on the momentum fraction variables $x_1$ or $x_2$, while 
$F_{j/{h_1}}$ and $F_{{\bar{k}}/{h_2}}$ in general depend on those kinematic
variables.\footnote{Here, and and throughout this work, the flavor dependence 
of the non-perturbative functions is ignored, as it is postulated in 
Ref.~\cite{CSS}.} 
The $\ln(Q^2/Q_0^2)$ dependence associated with the $F_1$ function was 
predicted by the renormalization group analysis \cite{CSS}. 
Furthermore, $F_1$ was shown
to be universal, and its leading behavior ($\sim b^2$) can be described by
renormalon physics \cite{Korchemsky-Sterman}. Various sets of fits to these
non-perturbative functions can be found in Refs.~\cite{Davies} and \cite
{Ladinsky-Yuan}.

In our numerical calculations, we use the Ladinsky-Yuan parametrization 
of the non-perturbative function (cf. Ref.~\cite{Ladinsky-Yuan}):
\begin{equation}
\widetilde{W}_{j\bar{k}}^{NP}(b,Q,Q_0,x_1,x_2)={\rm exp}\left[
-g_1b^2-g_2b^2\ln \left( {\frac Q{2Q_0}}\right) -g_1g_3b\ln {(100x_1x_2)}
\right] ,  \label{eq:WNPLY}
\end{equation}
where $g_1=0.11_{-0.03}^{+0.04}~{\rm GeV}^2$, $g_2=0.58_{-0.2}^{+0.1}~{\rm %
GeV}^2$, $g_3=-1.5_{-0.1}^{+0.1}~{\rm GeV}^{-1}$, and $Q_0=1.6~{\rm GeV}$.
(The value $b_{max}=0.5~{\rm GeV}^{-1}$ was used in determining the above $%
g_i$'s.) These values were fit for CTEQ2M PDF with the canonical choice of
the renormalization constants, i.e. $C_1=C_3=2e^{-\gamma _E}$ 
($\gamma _E$ is the Euler constant) and $C_2=1$. 
In principle, for a calculation using a different set of PDF, 
these non-perturbative parameters should be
refit using a data set that should include the recent high
statistics $Z^0$ data from the Tevatron.

In Eq.~(\ref{eq:ResFor}), $\widetilde{W}_{j{\bar{k}}}$ sums over the soft
gluon contributions that grow as 
$\alpha_S^n Q_T^{-2} \ln ^{2m -1}{(Q_T^2/Q^2)}$ ($1 \leq m \leq n$)
to all orders in $\alpha _S$. Contributions less singular
than those included in $\widetilde{W}_{j{\bar{k}}}$ should be calculated
order-by-order in $\alpha _S$ and included in the $Y$ term, introduced in
Eq.~(\ref{eq:ResFor}). This would in principle extend the applicability of
the CSS resummation formalism to all values of $Q_T$.\footnote{%
It is shown in Ref.~\cite{Balazs-Yuan} that since the $A$, $B$, $C$, and $Y$ 
functions are only calculated to some finite order in $\alpha _S$, 
the CSS resummed formula as described above will cease to be adequate 
when the value of $Q_T$ is in the vicinity of $Q$. Hence, in practice, one
has to switch from the resummed prediction to the fixed order perturbative
calculation as $Q_T\ge Q$.}
The $Y$ term, which is defined as the difference
between the fixed order perturbative contribution and those obtained by
expanding the perturbative part of $\widetilde{W}_{j{\bar{k}}}$ to the same
order, is given by 
\begin{eqnarray}
\ &&Y(Q_T,Q,x_1,x_2,\theta ,\phi ,C_4)=\int_{x_1}^1{\frac{d\xi _1}{\xi _1}}%
\int_{x_2}^1{\frac{d\xi _2}{\xi _2}}\sum_{n=1}^\infty \left[ {\frac{\alpha
_s(C_4Q)}\pi }\right] ^n  \nonumber \\
&&\ \times f_{a/h_1}(\xi _1,C_4Q)\,R_{ab}^{(n)}(Q_T,Q,\frac{x_1}{\xi _1},%
\frac{x_2}{\xi _2},\theta ,\phi )\,f_{b/h_2}(\xi _2,C_4Q),  
\end{eqnarray}
where the functions $R_{ab}^{(n)}$ contain contributions less singular than 
$\alpha_S^n Q_T^{-2} \ln ^{2m -1}{(Q_T^2/Q^2)}$ ($1 \leq m \leq n$)
as $Q_T\rightarrow 0$.
Their explicit expressions are summarized in Refs. 
\cite{Balazs-Qui-Yuan,Balazs-Yuan}.

\section{The Ratio $R_{CSS}$}\label{sec:RCSS}

\begin{figure*}[t]
\begin{center}
\begin{tabular}{cc}
\ifx\nopictures Y \else{ 
\epsfysize=6.0cm 
\epsffile{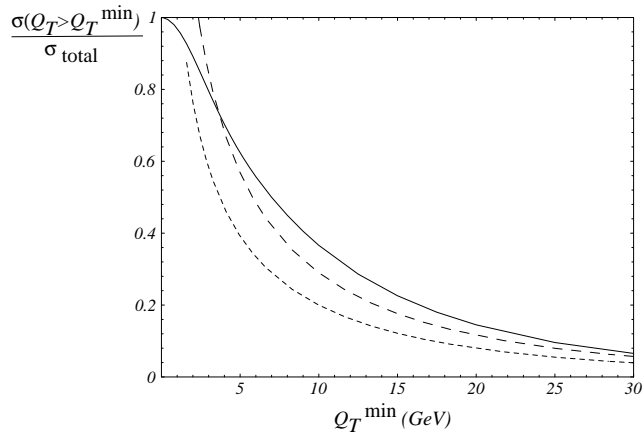} } 
\fi &  
\end{tabular}
\end{center}
\caption{ The ratio $R_{CSS}$ as a function of $Q_T^{\min}$ for $W^+$ bosons. 
The fixed order [${\cal O}(\alpha_S)$ short dashed, 
${\cal O}(\alpha_S^2)$ dashed] curves are ill-defined in the low $Q_T$ region.
The corresponding distributions for the $Z^0$ boson are indistinguishable from
those for the $W^\pm$ in this plot.}
\label{fig:Integrated}
\end{figure*}

In this work we propose to measure the ratio
$R_{CSS}(Q_T^{\min}) = \sigma (Q_T>Q_T^{\min })/\sigma _{Total}$ 
to distinguish the predictions of the
resummed, NLO and NNLO calculations. In Fig.~\ref{fig:Integrated} we show
the distributions of $R_{CSS}$, which is defined by 
\[
R_{CSS}(Q_T^{\min})\equiv \frac{\sigma (Q_T>Q_T^{\min })}{\sigma _{Total}}=
\frac 1{\sigma _{Total}}\int_{Q_T^{\min }}^{Q_T^{\max}}dQ_T\;\frac{d\sigma
(h_1h_2\rightarrow V)}{dQ_T}, 
\]
where $Q_T^{\max}$ is the largest $Q_T$ allowed by the phase space. In the
NLO calculation, $\sigma (Q_T>Q_T^{\min })$ grows without bound near $%
Q_T^{\min }=0$, as the result of the singular behavior $1/Q_T^2$ in the
matrix element. The NLO curve runs well under the resummed one in the 2 GeV $%
<Q_T^{\min }<$ 30 GeV region, and the $Q_T$ distributions from the NLO and
the resummed calculations have different shapes even in the region where $%
Q_T $ is of the order of 15 GeV.

With a large number of fully reconstructed $Z^0$ events at the Tevatron, one
should be able to use the data to clearly discriminate these two theoretical 
calculations. 
The experimental uncertainty in the total cross sections of the 
$W^\pm$ and $Z^0$ productions, based on 19.7 pb${}^{-1}$ CDF data, 
is in the ballpark of 5\% \cite{CDFTotal}. 
Fig.~\ref{fig:Integrated} shows that, 
in the 10 GeV $< Q_T^{\min} <$ 30 GeV region, 
even with this experimental precision we should see deviations 
between the experiment and the NLO predictions, in which
the effects of the multiple gluon radiation are not included.
In view of this result it is not surprising that the D0 analysis of the 
$\alpha_S$ measurement \cite{D0alphaS} based on the measurement of $\sigma
(W+1\;jet)/\sigma (W+0\;jet)$ does not support the NLO\ calculation. We expect
that if this measurement were performed demanding the transverse momentum
of the jet to be larger than about 50 GeV, at which scale the resummed and
NLO distributions cross (cf. Ref. \cite{Balazs-Yuan}), the NLO calculation
would adequately describe the data. 

To show that for $Q_T$ below 30 GeV the QCD multiple soft gluon radiation
is important to explain the D0 data \cite{D0alphaS}, we also include in Fig.~%
\ref{fig:Integrated} the prediction for the $R_{CSS}$ distribution at the order
of $\alpha _S^2$. As shown in the figure, the $\alpha _S^2$ curve is closer
to the resummed curve, which proves that for this range of $Q_T$ the soft
gluon effect included in the $\alpha _S^2$ calculation is important for
predicting the vector boson $Q_T$ distribution. In other words, in this
range of $Q_T$, it is more likely that soft gluons accompany the $W^{\pm }$ 
boson than just a single hard jet associated with the vector boson
production. For large $Q_T$, it becomes more likely to have hard jet(s)
produced with the vector boson.

The measurement of $R_{CSS}$ can also provide information
about the non-perturbative physics associated with the initial
state hadrons. As shown in Ref.~\cite{Ladinsky-Yuan}, the effect of the
non-perturbative physics on the $Q_T$ distributions of the $W^{\pm }$ and
$Z^0$ bosons produced at the Tevatron is important for $Q_T$ less than about
10 GeV. This is evident if one observes that different parametrizations of the
non-perturbative functions do not change the $Q_T$ distribution for $Q_T>10$
GeV, but can dramatically change its shape for $Q_T<10$ GeV. Since for
$W^{\pm }$ and $Z^0$ production, the $\ln (Q^2/Q_0^2)$ term is large,
the non-perturbative function, as defined in Eq.~(\ref{eq:Wnonpert}), 
is dominated by the $F_1(b)$ term which is expected to be universal for all 
Drell-Yan type processes and which is related to the physics of the renormalon 
\cite{Korchemsky-Sterman}. Hence, the measurement of $R_{CSS}$ can be used
to probe this part of non-perturbative physics for $Q_T<10$ GeV, in addition to 
probing the dynamics of multiple soft gluon radiation 
for 10 GeV $<Q_T<$ 40 GeV. It is therefore important to
measure $R_{CSS}$ at the Tevatron. With a large sample of $Z^0$ data at the
Run 2, it will be possible to determine the dominant non-perturbative function
which can then be used to determine the $W^{\pm }$ boson $Q_T$ distribution
to improve the accuracy of the $M_W$ and the charged lepton rapidity
asymmetry measurements.


\section{Conclusions}

In conclusion, the measurement of $R_{CSS}$  
provides an accurate test of the dynamics of the multiple soft gluon radiation 
predicted by QCD. 
With high enough luminosity, the NLO, NNLO, and resummed theoretical predictions 
can be distinguished by the $W^\pm$ and $Z^0$ data at the Tevatron. 
The comparison of the NNLO and resummed predictions shows that the soft gluon 
effect is important in the $Q_T<$ 30 GeV region.
Additionally, in the $Q_T<$ 10 GeV region, the $R_{CSS}$ measurement, 
using the recent $Z^0$ data at the Tevatron, can provide valuable information 
on the non-perturbative sector of the resummation formalism.
Therefore, a careful measurement of the $Q_T$ distribution of the vector bosons 
$W^\pm$ and $Z^0$ in the $Q_T< Q/2$ region at the Tevatron 
can further our knowledge 
of the perturbative dynamics and the non-perturbative domain of QCD. 


\acknowledgments
We thank R. Brock, G.A. Ladinsky, S. Mrenna, J.W. Qiu and W.K. Tung for
numerous discussions and suggestions, and to the CTEQ collaboration for
discussions on resummation and related topics. This work was supported in
part by NSF under grants PHY-9309902 and PHY-9507683.


\end{document}